\documentclass{article}
\usepackage[utf8]{inputenc}
\usepackage{url}
\usepackage{graphicx}
\usepackage{capt-of}

\title{Organising a Successful AI Online Conference: Lessons from SoCS 2020}
\author{Daniel Harabor$^1$ \and Mauro Vallati$^2$}
\date{
    $^1$Monash University, Australia\\%
    $^2$University of Huddersfield, UK
}

\begin{document}

\maketitle
\begin{abstract}
The 13th Symposium on Combinatorial Search (SoCS) was held May 26–28,
2020. Originally scheduled to take place in Vienna, Austria, the symposium
pivoted toward a fully online technical program in early March.  As an
in-person event SoCS offers participants a diverse array of scholarly
activities including technical talks (long and short), poster sessions,
plenary sessions, a community meeting and, new for 2020, a Master Class
tutorial program.  This paper describes challenges, approaches and
opportunities associated with adapting these many different activities to
the online setting. We consider issues such as scheduling, dissemination,
attendee interaction and community engagement before, during and after
the event. We report on the approaches taken by SoCS in each case, we give
a post-hoc analysis of their their effectiveness and we discuss how these
decisions continue to impact the SoCS community in the days after SoCS 2020.
This work will be of interest to organisers of similar conferences who
may be considering the switch to an online format.
\end{abstract}

\section{Introduction}

The Symposium on Combinatorial Search (SoCS) is an annual
meeting of AI researchers with an interest in the theory and practice
of symbolic state-space search.  Now in its 13th edition, SoCS 2020 took place
entirely online during May 26-28. The decision to hold SoCS as a virtual
conference was late-breaking and taken in response to the global
COVID-19 pandemic. It is a major departure, not only for the SoCS series
of symposia but also for the broader AI community, which values physical
meetings as the primary method for disseminating new scientific results.
On account of its early timing, SoCS-20 was regarded by some in the 
community as an experiment, ahead of other larger AI meetings that 
previously announced postponements but which did not yet commit to any 
specific format (e.g. CP, ICAPS, and IJCAI). 

In this paper we report results from the SoCS 2020 experiment, widely 
regarded amongst participants and in the community as a success. 
We describe the organisation of the symposium and our
decision-making process as regards different aspects of the 
technical program.
We were guided in this process by two important considerations:
(i) how to cater to the truly international community of SoCS attendees
and; (ii) how to maintain the ``SoCS community spirit'' where
attendees spend several days at a secluded location, 
working and socialising together at every opportunity. 

Beyond its immediate success, we also discuss how the now proven online
format adopted for SoCS 2020 is helping to shape the future of the community.
These issues include the need for broader digital footprint to help anchor
and grow the community online and the possibility for remote as well as
physical participation during future editions of the symposium.

\section{Background}


In the field of Artificial Intelligence, and in Computer Science more generally,
research moves extremely fast. Progress is driven by a combination of
extreme interest on the one hand and broad applicability on the other hand.
For this reason conference proceedings are primary method of scientific
communication. These areas are in contrast to other research disciplines,
such as Mathematics, Science and the Humanities, where journal publication
remains the preferred method of dissemination.

Since 2008 the Symposium on Combinatorial Search (SoCS) brings together a
diverse array of 50-60 AI researchers working on symbolic state-space search.
The latest results from this area appear in the SoCS technical
program and are preserved for future reference in archival proceedings.
Differentiating SoCS from other similar events is the fact that each year
the meeting takes place in a semi-remote ``retreat'' location which fosters
long periods of focused group discussion.  Because of its smaller size,
SoCS is usually planned to occur just before or just after another AI
meeting. For the 2020 edition, SoCS was organised as a co-located event
with the 17th International Conference on the Integration of Constraint
Programming, Artificial Intelligence and Operations Research (CPAIOR).
Both CPAIOR and SoCS were intended as physical meetings in Vienna, Austria,
with limited overlap to encourage cross-fertilisation.

With the rise of the COVID-19 global pandemic both SoCS and CPAIOR had
to pivot. The preferred option for SoCS was to retain the originally 
announced dates (May 26-28) and to migrate the entire technical program 
to an online setting. 
SoCS-20 was not the first conference to grapple with these challenges.
AAMAS-20, a related but substantially larger sister event, also pivoted in 
towards fully online and took place May 9-13, immediately preceding SoCS-20.
Other recent events include (but are not limited to): 
the International Conference on Performance Engineering 
(ICPE-20, April 20-24)~\cite{iosup2020flexibility}, 
the International Conference on Extending Database
Technology 
(EDBT, 30 March - 2 April)~\cite{bonifati2020holding}, 
Neuromatch (March 30-31, 2020)~\cite{achakulvisut2020point}
and the Photonics Online Meetup (POM-20, January 13, 2020)~\cite{pom2020}.

\section{Decisions to take}
The originally announced format for SoCS 2020 featured a diverse number of activities:
paper talks (short and long), invited plenaries, poster sessions,
a series of Master Classes (i.e. tutorials) and a community meeting.
In this section we discuss the main considerations and decisions taken 
in order to transfer this technical program to a fully online setting.


\subsection{Format for Technical Talks}
\label{ref:talks}

One of the pivotal aspects to consider when organising a conference is how talks will be delivered. 
In an online setting, talks can be delivered either live, using tools for online meetings such as \texttt{Zoom}, \texttt{Google Meet}, or \texttt{Microsoft Teams}, or under the form of pre-recorded videos. Both approaches have of course advantages and disadvantages, that we considered for SoCS 2020 and can be summarised as follows.

\paragraph{Live talks.}
Similar to an in-person meeting in terms of organisation and interaction,
the live talk format is well understood by speakers and organisers. It has
the least overhead in terms of setup costs (choosing a platform, scheduling
a time) and it allows speakers to work on the slides up to the time of
their talk.  One of the main disadvantages is scheduling: speakers must
be available at fixed times and in a suitable environment for presenting.
Another disadvantage is that the quality of the presentation depends on the
quality of the network connection, not only for the speaker but also for
participants.  It has been documented, for example, that during the COVID-19
lock-down period Internet usage dramatically increases~\cite{c2020impact}
which affects quality and reliability of individual connections.


\paragraph{Pre-recorded talks.} 
In this format speakers record and submit their talk well in advance. 
The quality of the content is carefully controlled by the speaker
and the quality of the delivery is guaranteed once the video is 
shared and downloaded. 
The main disadvantage of this format is an increased workload for 
organisers. Detailed instructions have to be provided
for speakers before the event.
Submitted videos must then checked 
(for video and audio quality, for adherence to time limits)
and possibly post-processed, such as into a streaming session.
The timing and release of videos is another area that requires 
careful consideration so that each talk can receive the attention
of the community.

%
%
%
%
%

\paragraph{Impact on Q\&A.}
The format of talks strongly influences the type of interactions
possible between speakers and participants.  Live talks must be carefully
managed and questions can only be taken at the end. Depending on the
timing of the speaker, and the constraints of the schedule, discussion can
even be cut short so that the next presentation can begin. By comparison,
pre-recorded talks have the advantage that discussion can take place during
the premiere time of each video, which gives participants and speakers
more time to interact.

%

For SoCS 2020 we carefully considered the pros and cons of each format
and opted for pre-recorded talks.  A few days after the acceptance
notification, authors were provided with detailed instructions for
recording their talks, using tools such as \texttt{Screencast-o-matic},
\texttt{Kazam Screencaster} and \texttt{OBS Studio}. We requested each
talk begin with a title card\footnote{\url{https://tinyurl.com/txdpkhm}}
showing photos of the speaker and possibly the authors. For keeping the
talks engaging, we recommended strategies such as colourful slides and
animations and using a picture-in-picture view to show the presenter
alongside the content.  We required authors to submit their videos three
weeks before the conference. This allowed for some delays in the process
(to be expected, given the exceptional circumstances) and enough time for
us, the organisers, to check videos and prepare them.
For Q\&A, we used a combination of synchronous live chat and an
asynchronous discussion forum (see Section~\ref{sec:interaction}).

\subsection{Posters}
Each year SoCS receives a substantial number of extended abstracts 
which are presented during a poster session. 
These sessions provide participants an opportunity to browse a large
number of works and to have longer 1-1 discussions.
There is no clear online equivalent for these types of interactions. 
Recent online conferences such as EDBT-20 ~\cite{bonifati2020holding}
suggest converting poster sessions into short talks, organised
into a dedicated session without Q\&A.
At POM-20~\cite{pom2020} posters were presented as a deck of 
4 slides, each announced and discussed on Twitter. 

For SoCS-20, we developed a specific ``Micro Talk'' format for
poster presentations. Each talk was limited to 5 minutes and a
maximum of 3 slides (not including title card). These talks
were mixed into regular sessions. During and immediately after 
the premiere of each Micro Talk there was a live Q\&A 
with opportunities for asynchronous forum discussion thereafter.


\subsection{Attendance and Registration}\label{sec:registration}

Online conferences have many benefits compared to in-person meetings and they
are are well positioned for attracting a wider audience.  For participants,
travel costs are eliminated, registration costs are reduced and substantial
amounts of time are saved.  For organisers, online conferences are simpler to
plan and less expensive: many aspects such as catering, rooms, receptions,
badges, welcome packs, etc. are avoided. Another benefit, for attendees and
organisers, is reducing the environmental impact compared to conventional
meetings~\cite{higham20}.
Some new complications arise, such as hosting fees and licences, but  
these overheads tend to be significantly smaller than those for an
in-person meeting.  The costs can be recovered by asking attendees to pay
a registration fee for online attendance however this can act as a barrier
to wider participation.

For SoCS 2020 we decided to charge a registration fee only to one author
per accepted paper, which covers the cost of the proceedings. Beyond that
participation was free for everyone. We asked authors, speakers,
and interested participants to register for the SoCS forum, a dedicated 
bulletin board which we used as a channel for communication and further 
announcements. The talks of the authors were free to watch, even anonymously,
being streamed directly to YouTube at fixed premiere times.
Following their initial release, videos became freely
available for viewing on-demand\footnote{Links to the session
videos and individual talks appear on the SoCS-20 website:
\mbox{\url{http://socs20.search-conference.org/main.php?page=program}}}
and we plan that they remain so in the foreseeable future.


\subsection{Scheduling and Program}\label{sec:program}

While the scheduling and the organisation of a conference program is always
a critical task, it becomes particularly challenging for online events. In
the case of pre-recorded videos, one tempting option could be to release
all talks at the same time. Indeed this option was explored at AAMAS 2020, 
with only live keynotes and plenaries being scheduled at fixed times. 

In the case of SoCS, we felt that releasing all videos simultaneously would 
undermine the spirit of the conference, which intends to bring together
a tightly knit community for focused interaction.
Instead, we opted for a conventional (to AI conferences) format, with 
talks being organised thematically into sessions. 
Each session ran for approximately 60-90 minutes and was scheduled for 
release on YouTube using the \textit{Premiere} feature. 
In this setup videos are played one after the other at fixed times, with 
some minutes of intermission in between. 
We found that sessions draw larger audiences, which means each talk has a 
chance to be in the spotlight. Also that sessions encourage the community 
to gather at fixed times, which results in more vibrant Q\&A with speakers 
and more robust interactions among participants.
We scheduled 5-minutes breaks between talks in each session. These breaks
signal the end of the live Q\&A session (which begins while the video plays) 
and they provide a well-delimited time window for the community to re-synchronise.
Longer breaks, between sessions, simulate the lunchtimes and coffee breaks of a
conventional meeting. They allow for longer discussions and for social 
gatherings. We observed that many participants at SoCS-20 formed social 
circles during longer breaks, and many participants met virtually in the evenings, 
after the end of the daily program. 

This type of fixed scheduling requires one to select a reference time
zone. However, attendees significantly out-of-sync with the reference time-zone 
will likely be unable to join for synchronous interaction. 
After careful consideration we opted for Central European Time (CET). 
This was done for a number of reasons: first, because the in-person conference
was supposed to happen in Austria; second, because Europe is the region
where the largest number of community members live; third, because 
CET provides a middle ground between people living
in the Americas, and people living in Asia and Australia. To minimise the
discomfort for them, where possible, talks involving authors from Asia
and Australia were scheduled in the morning, while talks involving authors
from Americas where scheduled in the late afternoon.


\subsection{Interaction between Participants}\label{sec:interaction}

A major aspect of any AI conference is the interaction between
participants. When moving to a virtual setting it becomes crucially important
to select the right tools and to leverage online advantages.  At one end of
the spectrum of possibilities we have asynchronous interaction. This method
of communication is typified by email correspondence and discussion forums.
It does not require participants to be active at the same time and it
allows for more elaborate and articulated exchanges. At the other end of
the spectrum is synchronous interaction. This approach captures the spirit
of face to face discussion: it is fast moving and requires participants
to be online at the same time.  For SoCS 2020, we decided to support both
types of interaction.

For asynchronous interaction we used a dedicated
forum\footnote{\url{http://forum.search-conference.org/}}. 
We created discussion threads for each keynote and session.
We also posted announcements to the forum and we used it to provide 
instructions and information ahead of the symposium, such as advice 
for recording videos and instructions for participating. The forum 
was free, but registration was required to minimise the risk of
trolls~\cite{BUCKELS20149}. Trolls represent a critical risk for free online
events, as they can disrupt any meaningful interaction and communication.

Synchronous interaction was achieved using the Discord
platform\footnote{\url{https://discord.com/}}. Discord supports video calls,
voice chat, and text chat. Instructions on how to access the dedicated
SoCS Discord channel, and how to use Discord, were provided via the forum
--- again to reduce the possibility of trolls. Authors of accepted papers
were also asked to make themselves available for live Q\&A on Discord at 
the time when their talk was scheduled for premiere.
During longer breaks, public and private group discussions sprang spontaneously,
as happens during breaks at in-person conferences.


\subsection{Moderation}

Moderation requires significant effort for in-person AI conferences: a large
number of session chairs and helpers are usually needed to make sure that,
for instance,  the schedule is followed, discussions do not degenerate, and
that participants are behaving according to an understood code of conduct.
Our experience at SoCS-20 is that moderation in virtual setting is important
but much less demanding. 
With regards to the forum, we used only very lightweight moderation: mostly for 
maintaining the organisation and structure of the content.  For the
Discord, we found that two individuals (the organisers)  were enough to
to moderate all the sessions. 
In all cases we relied on the ICAPS code of conduct\footnote{\url{http://www.icaps-conference.org/index.php/Main/CodeOfConduct}}
to make clear the expected standard of all interactions.

Regarding Discord, we did not observe any abusive or aggressive
behaviour. Instead, we were pleasantly surprised by the fact that questions
were more elaborated and more ``friendly'' in tone might be expected at
an in-person meeting. We assumed it was because with text it is hard
to identify tones and that participants spent additional time to make
sure their questions and their answers read well. We also noticed that
conventions quickly emerged in the community: ways for ``clapping''
at the end of talks, for instance, and the best way to ask questions
(by using the mentioning features of Discord) to make sure that they
were noticed. Notably, such conventions evolved during the time of the
conference, and were widely and promptly adopted by all the participants,
with no enforcement from the moderators.
In some cases Q\&A discussions went on for too long and overlapped with
the start of the next talk. In such cases we asked the involved participants 
to move the discussion to a dedicated chat or to the forum. 
This is a very nice plus of
online conferences, as one group of attendees can continue to interact 
while others can follow sessions without being disturbed.


\subsection{Keeping Records}

Discussion and Q\&A for in-person conferences are highly valuable aspects,
as they can lead to new collaborations and highlight potential developments
of the presented works. Unfortunately, they are ephemeral in nature, as it
is almost impossible to keep accurate public records during an in-person meeting. This
is not the case for online conferences, where textual discussions and Q\&A
can be easily recorded for posterity. It is of course important to decide
how to structure such minutes so that they are accessible and searchable
by the community at a later point in time.

For SoCS 2020 we kept track of all the discussions that happened during
Q\&A sessions. The forum provided the ideal platform for recording such
discussions, and for structuring them. In particular, one thread per
session was created to maximise clarity and to make it easier to look
at them. The main point was to make sure that people that did not attend the
specific session were provided with all the relevant and interesting discussion
from the session. Being asynchronous, the forum allows these discussions to
continue even after the conference is over, and to provide links to relevant 
papers, tools, or websites mentioned during the meeting.

Beside discussions, videos of the talks are another way of documenting
the conference, and they can be likewise organised and stored. For SoCS 2020, we
created playlists in a dedicated YouTube channel. The description of
each video specifies its content, and refers to the conference forum
for additional information. The complete playlist is available at
\url{http://shorturl.at/jwIY5}.


\section{Evaluation}
\label{sec:eval}


To evaluate the success of SoCS 2020 we examine participation
rate, which we define as the number of registered participants per
accepted paper. As is typical for the field, SoCS requires at least
one author per paper to register for the conference and to present 
the work as part of the technical program. In other words, the minimum 
participation rate expected in any given year is 1.0.
The underlying hypothesis is that higher rates of participation indicate the technical program has 
attracted the attention of a broader community.

Table~\ref{tab:ppr} compares SoCS 2020 with the previous five editions 
(2015-2019), each of which were held as in-person meetings and
without any online component. 
The data show that, as a physical event, SoCS is typically
attended only by authors of accepted works.
Higher rates of participation can be observed in years where the conference
is held in a location with a strong community presence (Israel, 2015)
or when the conference is co-located and concurrent with major
AI meetings (IJCAI, 2018).
SoCS 2020 has by far the highest recorded participation rate,
despite a smaller technical program. 
We attribute these gains to the online format and to the low cost.
With free registration and no travel requirements multiple authors 
per paper can register for the conference.  
There were 140 registrations on the SoCS forum and 107 unique users  
subsequently logging into the SoCS Discord server for live discussions
(the address of our Discord server was made available only to forum members).

\begin{table}[t]
\begin{tabular}{|l|l|c|c|c|}

\hline
{\bf Year} & {\bf Location} & {\bf Accepted Papers} & {\bf Participants}  & {\bf Ratio} \\ \hline
2015 & Ein Gedi (Israel) & 44 & 65 & 1.4 \\ 
2016 & Tarrytown (USA) & 41 & 41 &  1.0 \\ 
2017 & Pittsburgh (USA) & 49 & 54 & 1.1 \\
2018 & Stockholm (Sweden) & 27 & 65 & 2.4 \\
2019 & Napa (USA) & 46 & 50 & 1.1 \\ 
2020 & Online & 27 &  140 & 5.2 \\ \hline

\end{tabular}
\label{tab:ppr}
\caption{Acceptance and participation rates for SoCS conferences from 2015 to 2020.
The number of papers is taken by reference to published proceedings. 
The number of participants is taken from data collected by previous organisers 
(2015-2019) and by counting user registrations on the SoCS 
forum (2020).}
\end{table}

Although free, user registration can act as a disincentive for
persons otherwise interested in online content~\cite{li2014drives}.
To mitigate this issue every session at SoCS 2020 was premiered 
on YouTube and made available afterwards, as detailed in Sections \ref{ref:talks} and \ref{sec:program}, so anyone with Internet access can enjoy them. 
User registration was therefore necessary only for Q\&A with the 
speakers and for interacting with other participants from the SoCS community, as described in Sections \ref{sec:registration} and \ref{sec:interaction}. 

Figures~\ref{fig:views} and \ref{fig:views2} show statistics gathered from
YouTube regarding views and viewers.
We focus on the 3 days of the conference and the week 
immediately after.

\vspace{2em}
\begin{minipage}[t!]{0.45\textwidth}
	\includegraphics[width=\textwidth]{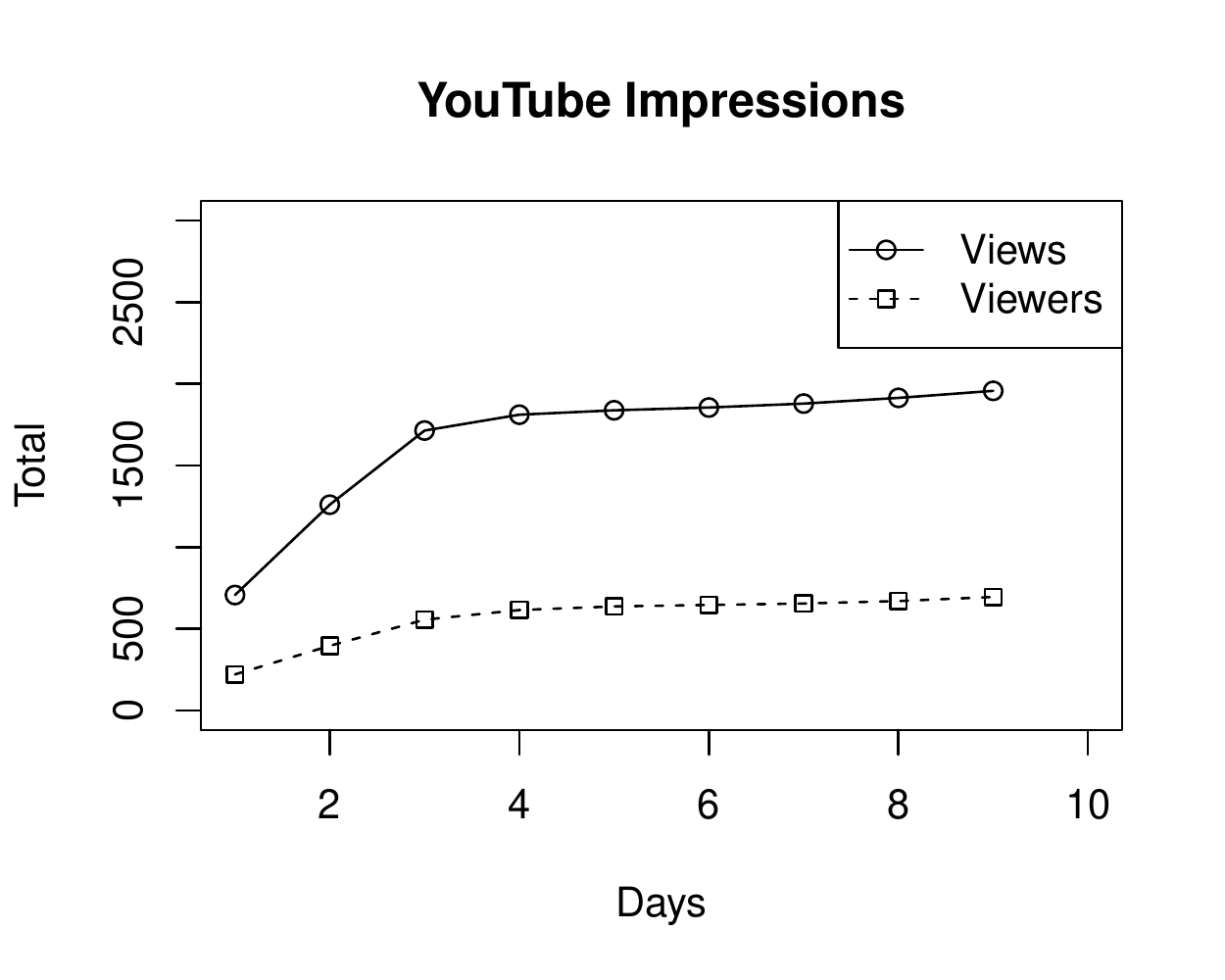}
	\captionof{figure}{\label{fig:views}Cumulative views and viewers, according to YouTube.}
\end{minipage}
\hfill
\begin{minipage}[t!]{0.45\textwidth}
	\includegraphics[width=\textwidth]{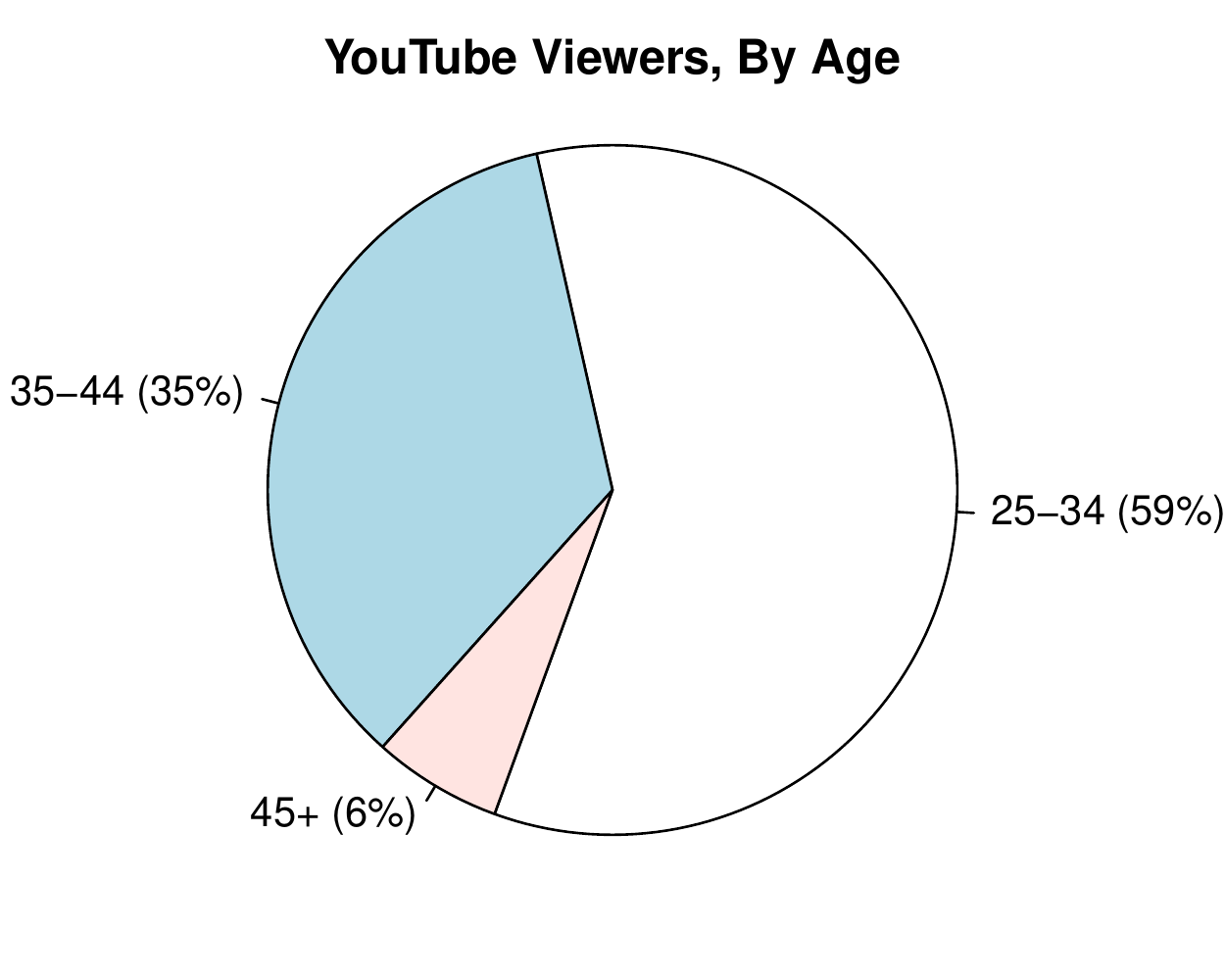}
	\captionof{figure}{\label{fig:views2}Viewers by age, according to YouTube. }
\end{minipage}
\vspace{2em}

\noindent
YouTube reports 1957 unique views over this period, 
approximately two thirds of which occur in the first three days.
There were 469 unique viewers in total, or approximately three times 
higher than the number of registered participants.
Moreover, with all presentations being available online and in
perpetuity, there exist further opportunities for interested
persons to find and engage with the technical material. Here we focus on
that period of time because, in our opinion, it provides the best angle to
analyse the conference success. Videos on YouTube will still be watched in
the coming months, but that is a different kind of evaluation, that looks
more into the impact of the topic for the wider --and potentially
non-academic-- community.



\section{Discussion and Conclusions}

SoCS 2020 received an extremely strong positive response from attendees and 
from the the Search community more generally. We feel confident concluding
that moving the conference to a fully online format, despite some risks and
uncertainties, was ultimately the right decision.
The core principles of our approach can be summarised as follows:
\begin{itemize}
\item {\bf Pre-recorded videos}. 
This approach allows speakers to carefully manage the quality 
of their material and delivery, while avoiding all technical 
issues typically associated with live presentations.

\item {\bf Streaming sessions}.
This approach allows the community to meet online at times announced
well in advance. Because sessions attract larger audiences, every video has an opportunity 
to be in the spotlight.  After the session, videos are available for viewing on-demand.

\item {\bf Live Q\&A}.
This approach has the advantages that discussions can take
place during the video premiere, instead of only at the end 
as with a conventional format. We found that live chat 
works well for a smaller community such as SoCS but it also
has the potential to scale to larger events, where moderators can
relay questions to the speaker. 

\item {\bf Community hub}. 
We used a discussion forum for asynchronous discussion where
participants and speakers can engage after a video premiere.
The forum also served to coordinate the conference and for keeping
a record of the meeting, with live discussions being summarised there.
\end{itemize} 

\noindent
By the time of the community meeting, which typically concludes every 
SoCS event, it was clear the online format had become a proven success.
Among the many issues arising at the meeting was whether future editions
of SoCS should continue as online meetings or at least retain some
online aspects. Among the identified advantages are higher participation 
rates, reduced costs, and a much smaller environmental 
impact~\cite{higham20,pacchioni2020virtual}.
One possibility for a mixed format is to introduce a pre-recorded 
``micro talk'' which can serve as an advertisement for a longer in-person
event, but can still give an overview of the paper to people that are not able to attend the in-presence event. Another even more blended possibility is the addition of a ``virtual day'' which 
could precede the in-person meeting and include additional activities
such as a Doctoral Consortium or further Master Class talks. 

Other innovations from SoCS 2020 are already having an impact on the community. 
The discussion forum, for example, has been adopted as a general hub for the 
discussion of the SoCS series of symposia and for search-related topics more 
generally. 
Videos uploaded to the SoCS YouTube channel will form part of an
upcoming library intended to bring students and newcomers up to the moment 
with research directions in the area.



\bibliographystyle{plain}
\bibliography{references}

\end{document}